\documentclass[%
 reprint,
 amsmath,amssymb,
 aps,
]{revtex4-2}

\usepackage{graphicx}
\usepackage{dcolumn}
\usepackage{xcolor}
\usepackage{bm}
\usepackage{hyperref}
\usepackage[mathlines]{lineno}

\begin{document}

\preprint{APS/123-QED}

\title{Aging effects in Schelling Segregation Model}

\author{David Abella}
 \altaffiliation[]{david@ifisc.uib-csic.es}
\author{Maxi San Miguel}%
\author{Jos\'e J. Ramasco}%
\affiliation{Instituto de F\'{\i}sica Interdisciplinar y Sistemas Complejos IFISC (CSIC-UIB), Campus UIB, 07122 Palma de Mallorca, Spain}%

\date{\today}

\begin{abstract}

The Schelling model has become a paradigm in social sciences to explain the emerge of residential spatial segregation even in the presence of high tolerance to mixed neighborhoods by the side of citizens. In particular, we consider a noisy constrained version of the Schelling model, in which agents maximize its satisfaction, related to the composition of the local neighborhood, by infinite-range movements towards satisfying vacancies. We add to it an aging effect by making the probability of agents to move inversely proportional to the time they have been satisfied in their present location. This mechanism simulates the development of an emotional attachment to a location where an agent has been satisfied for a while. The introduction of aging has several major impacts on the model statics and dynamics: the phase transition between a segregated and a mixed phase of the original model disappears, and we observe segregated states with high level of agent satisfaction even for high values of the tolerance. In addition, the new segregated phase is dynamically characterized by a slow power-law coarsening process and by a glassy-like dynamics in which the asymptotic time translational invariance is broken.
\end{abstract}

\maketitle

\section*{Introduction}

\smallskip

Thomas Schelling introduced a simple segregation model   \cite{schelling-1969,Schelling,schellingbook,hegselmann-2017} in which agents of two colors are distributed randomly on a chess-board, leaving some locations free. Agents are unsatisfied if more than a half of the eight nearest neighbors have different color. Randomly, the unsatisfied agents will move to available satisfying locations of the neighborhood. This model has had a very significant impact for several reasons: The "hand-made" simulations performed by T. Schelling by moving pawns on a chessboard are an early precedent of the use of  agent-based simulations in Social Sciences. It is also one of the first social models to show emergent behavior as a result of simple interactions among agents, a characteristic of complex systems. A robust result of the model is that segregation occurs even when individuals have a very mild preference for neighbors of their own type, so that collective behavior is not to be understood in terms of individual intentions. In addition, the model introduced the concept of behavioral threshold that inspired a number of other models of collective social behavior \cite{granovetter}. But still currently, Schelling´s model is at the basis of fundamental studies of the micro-macro paradigm  in Social Sciences \cite{grauwin-2009}, while it continues to have important implications for social and economic policies addressing the urban segregation problem \cite{clark-1991,Sassen,Clark,lamanna-2018}.

As a result of the notable implications of this model and the robustness of the emerging segregation, there exists a vast literature around Schelling's results. Many variants of the original Schelling model have been reported modifying the rules that govern the dynamics, the satisfaction condition, or including other mechanisms, network effects, or specific applications \cite{Vinkovic,stauffer-2007,Dall_Asta_2008,gracia-lazaro-2009,Gauvin_2009,Gauvin_2010,domic-2011,henry-2011,unified,Interfacial_roughening,stauffer-2013,lenormand-2015,barmpalias-2018,jensen-2018,holden-2019,sert-2020,agarwal-2020,vieira-2020,ortega-2021,ortega-2021.2}. In particular, the Schelling model has been studied from a Statistical Physics point of view due to its close relation to different forms of Kinetic Ising-like models \cite{stauffer-2007,stauffer-2013}, and also addressing general questions of clustering and domain growth phenomena, as well as for the existence of phase transitions from segregated to non-segregated phases. For example, the relation with phase separation in binary mixtures has been considered \cite{Dall_Asta_2008,Vinkovic}, as well as the connection with the phase diagram of spin-1 Hamiltonians \cite{BEG,BlumeCapel,Gauvin_2009,Gauvin_2010}.
In this context a useful classification of models is to distinguish between two possible types of dynamics \cite{Dall_Asta_2008}: "constrained", where agents just move to satisfying vacancies (if possible) and "unconstrained",  where agents motion do not prevent them to remain unsatisfied. In addition, the motion can be short range (only to neighboring sites as in the original model) or long range. Constrained motion is "solid-like" generally leading to frozen small clusters, while unconstrained motion is "liquid-like" allowing for large growing clusters \cite{Vinkovic}. Including the motion of satisfied agents leads to a noisy effect playing the role of temperature in a statistical physics approach. 

Our goal is to characterize how "aging" modifies the segregation dynamics of the Schelling model. Aging takes into account how the persistence of an agent in a given state modifies the transition rate to a different state \cite{fernandez-gracia-2011,perez-2016,boguna-2014}. This concept of aging, or inertia \cite{Stark2008}, constrains the transitions in a way that the longer an agent remains in a given state, the smaller is the probability to change it. This rate dependence on the persistence times accounts for the observation that human interactions do not occur at a constant rate. They rather show  a bursty character with a non-Poissonian inter-event time distribution \cite{barabasi-2005,moro,oriol,rybski-2012,zignani-2016,kumar-2020}. However, most social simulations, including simulations of variants of the Schelling model, implicitly assume a constant rate of interactions or state updating. Nevertheless, aging has been already shown to modify social dynamics very significantly. For example, in opinion dynamics, aging is able to produce coarsening towards a consensus state in the voter model \cite{fernandez-gracia-2011,peralta-2020} or to induce a continuous phase transitions in the noisy voter model \cite{artime-2018}. 

In this paper, aging is introduced in the Schelling model by considering that agents are less prone to change their location as they get older in a satisfying place. In other words, aging is introduced giving a smaller probability for the  "moving-out" of satisfied agents the longer they have remained in a satisfying neighborhood. We implement this aging mechanism in the long range noisy constrained version of the Schelling Model \cite{Gauvin_2009}, for which a detailed phase diagram was reported. We study how this phase diagram is modified by the aging mechanism, finding that aging inhibits a segregated-mixed phase transition. This implies that aging favors segregation, a counter-intuitive result. We also describe the coarsening dynamics in the segregated phase and associated autocorrelations \cite{puri-2004}, showing that aging gives rise to a slower coarsening and to a glassy type-dynamics with breaking of the time-translational invariance.



\section*{Methods} 

\subsection*{Model}

The model considered in this work is a variant of the noisy constrained Schelling model \cite{Gauvin_2009} in which we explicitly include aging effects. For simplicity, we refer to this variant as Schelling model during the rest of the paper to compare with the model presented here: the Schelling model with aging. For both, the system is established on a $L \times L$ Moore lattice with $8$ neighbors per site and periodic boundary conditions, where agents of two kinds (representing, for instance, wealth levels, race, language, etc) occupy the sites. There are also empty sites (vacancies), to where agents can move depending on their state and on the vacancy neighborhood. The condition of each site $i$ of the lattice will be described with a variable  $\sigma_i$ that takes three possible values: $\sigma_i = \pm 1$ for the two kinds of agents and $\sigma_i = 0$ for vacancies. In addition, depending on the local environment, agents can be in two states: satisfied or unsatisfied. In our case, agents are satisfied if their neighborhood is constituted by a fraction of unlike agents lower than a fixed homogeneous parameter $T$. Otherwise, they are unsatisfied. Therefore, this control parameter $T$ is a measure of how tolerant the population of the system is. We also need a non-zero vacancy density, $\rho_v > 0$, for agents to change their location. This $\rho_v$ is understood as an extra parameter of the model. The initial configuration is built by randomly distributing the agents ($N_{\rm agents} = L^2 \, (1 - \rho_v)$). We always consider one half of agents of each kind.

In the Schelling model considered, an agent chosen by chance moves to a random satisfying vacancy (if any exists) independently of his/her initial state and of the distance. This process is repeated until the system reaches a stationary state. The movement of unsatisfied agents behaves as a driver for the system dynamics, while the motion of satisfied agents plays the role of a noise. When tolerance $T$ becomes larger, more satisfying vacancies are present in the system and the noise consequently increases. 

The aging mechanism in our model is introduced by considering an activation probability of the agents inversely proportional to the time spent at the current state \cite{artime-2018}. This methodology was proposed to mimic the power-law like inter-event time distributions observed in real-world social systems \cite{barabasi-2005,fernandez-gracia-2011}. If an agent $j$ is initially satisfied in her neighborhood, the internal time is set $\tau_j = 0$. Then, in every time step a randomly chosen agent $j$ follows different rules depending whether she is originally satisfied or not. If unsatisfied, $j$ moves to any random satisfying vacancy of the system. Otherwise, she moves to another satisfying vacancy with an activation probability $p_j = 1 / (\tau_j + 2)$. In both cases, if no vacancy has a satisfying neighborhood, the agent $j$ remains in the initial site. As before, these rules are iterated until the system reaches a stationary state (if possible). The time is counted in Monte-Carlo steps; after $N_{\rm agents}$ iterations, the internal time increases for all satisfied agents in one unit, $\tau_j \to \tau_j + 1$. As for the Schelling model, there is a noise effect associated to the motion of satisfied agents. In this case, the intensity of this noise is related not only to the tolerance parameter $T$, but to the presence of aging as well. In fact, aging introduces more constraints to the movements and contributes to decrease the noise.  

Given the number of neighbors available in the Moore lattice, numerical simulations are only performed for a finite set of meaningful tolerance values: $\{1/8,1/7,1/6, \cdots ,6/7,7/8 \}$. During all our analysis, we focus on the low vacancy density region of the phase diagram. In this region, there is an even smaller number of meaningful $T$ values $\{1/8,2/8,...,7/8\}$, because the majority of agents do not see vacancies in their surroundings.

\begin{figure*}
\centering
\includegraphics[width=0.8\linewidth]{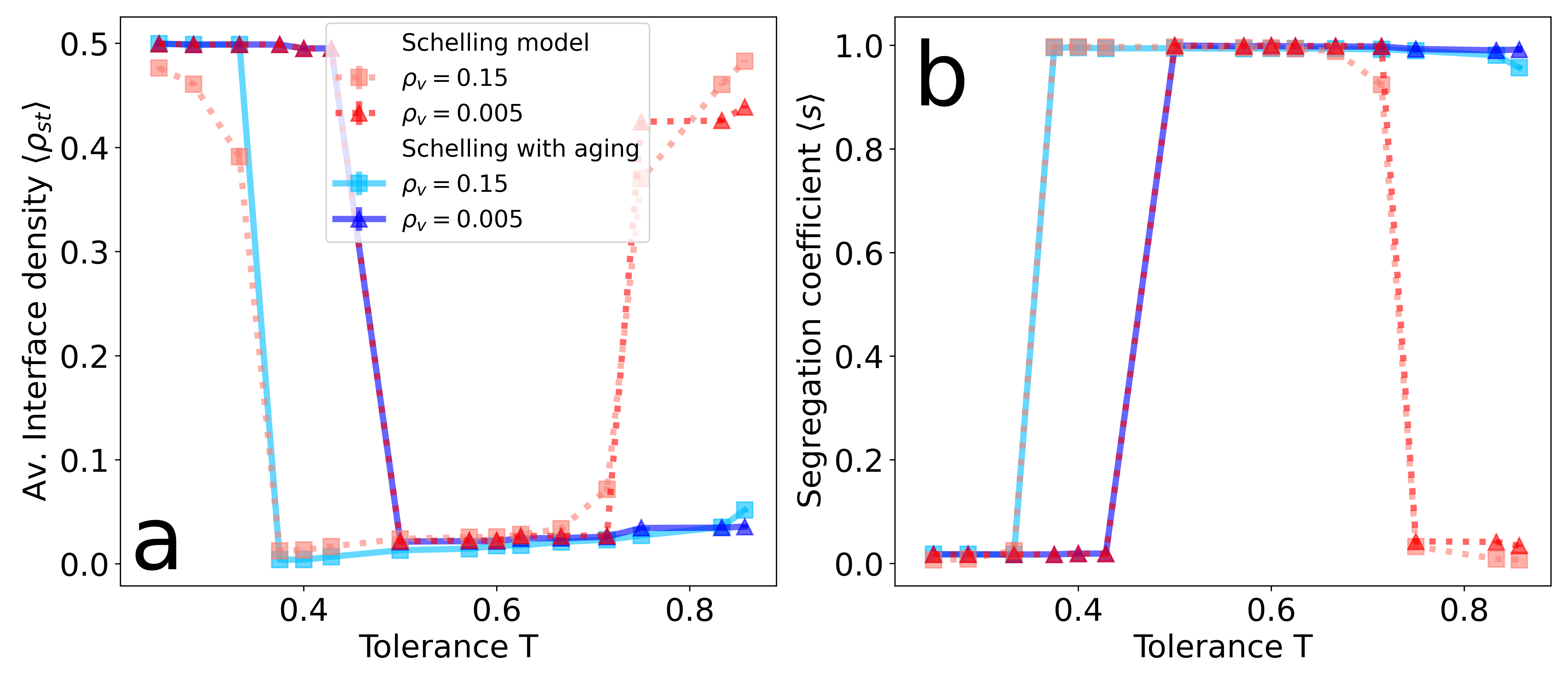}
\caption{Average interface density $\langle\rho_{\rm{st}} \rangle$ (\textbf{a}) and segregation coefficient $\langle s \rangle$ (\textbf{b}) at the stationary regime as a function of the tolerance parameter $T$ for two values of the vacancy density $\rho_{v} = 0.5\%$ and $15\%$. Results are shown for both the Schelling model and the variant with aging introduced in this paper. Simulations are performed on a $80\times 80$ lattice and averaged over $5 \cdot 10^{4}$ realisations.}
\label{Fig1}
\end{figure*}

\subsection*{Metrics of segregation}

Many metrics have been introduced in the literature to discern if the final state is segregated or not \cite{Gauvin_2009,lenormand-2015,randomwalks,urban}. The number of clusters is known to be directly related with the segregation, because a high presence of small clusters indicates a mixing between agents. As for the Schelling model\cite{Gauvin_2009}, we compute the following metric related to the second moment of the cluster size distribution:
\begin{equation}
s = \frac{2}{\left(L^{2} \, (1-\rho_v)\right)^{2}} \sum_{\{c\}} n_{c}^{2} ,
\end{equation}
where the index of the sum $c$ runs over all the clusters $\{c\}$ and $n_c$ is the number of agents in cluster $c$. The average of $s$ over realizations after reaching a stationary state is defined as the segregation coefficient $\langle s \rangle$. This metric is bounded between 0 and 1: $\langle s \rangle \to 1$ if there are only 2 equally-sized clusters, and $\langle s \rangle \to 0$ if the number of clusters tends to the number of agents. The cluster detection is performed using the Hoshen-Kopelman algorithm \cite{HoKo}.

Another metric of segregation is the interface density defined as the fraction of links connecting agents of different kind. The calculation is done in two steps: estimating the interface density for each agent $j$, $\rho_j$, and then the average over all the agents $\rho$:

\begin{equation}
    \rho_j = \frac{1}{2} \, \left( 1 - \frac{\sigma_j \,  \sum_{k \in \Omega_j} \sigma_k}{\sum_{k \in \Omega_j} \sigma_k^2 } \right) \quad \rm{and} \quad \rho = \frac{1}{N_{\rm agents}} \sum_{j = 1}^{N_{\rm agents}} \rho_j ,
\end{equation}

where the indices $k$ run over the neighborhood of agent $j$, $\Omega_j$. If an agent $j$ is surrounded only by vacant sites, we define by convention $\rho_j = 0$. Performing a realization average of $\rho$, we obtain the average interface density $\langle \rho \rangle$ in the stationary state is denoted as $\langle\rho_{\rm{st}} \rangle$. The evolution of this metric allows us to study the coarsening process.

\section*{Results} 

\subsection*{Phase diagram}

To discuss the phase diagram of our model, we focus on the region of parameters with a vacancy density $\rho_v < 50 \%$ to avoid diluted states with a majority of vacancies.
For this region, the Schelling model presents 3 different phases \cite{Gauvin_2009}: frozen, segregated and mixed. For low tolerance values, the system freezes in a disordered state given that there are no satisfying vacancies for any kind of agents. Increasing tolerance, the system undergoes a transition towards a segregated state, which is characterized by a 2-clusters dynamical final state. Finally, for high values of $T$, after another transition, we find a dynamical disordered (mixed) state, in which a vast majority of vacancies is satisfying for both kinds of agents and small clusters are continuously created and annihilated.

These three phases are characterized by measuring the segregation coefficient $\langle s \rangle$ and the average interface density $\langle\rho_{\rm{st}} \rangle$ at the final state. The results for the original model are depicted as a function of the tolerance $T$ in Fig. \ref{Fig1}a for the interface density and in Fig. \ref{Fig1}b for the segregation coefficient. At low values of T, both indicators show a disordered state that falls in the frozen phase.  We also observe a dependence of the transition point with the vacancy density. On the other hand, for high $T$ values, the transition point between segregated and mixed states has no dependence with the parameter $\rho_v$. Notice that mixed and frozen states present a very similar value of $\langle s \rangle$ but can be differentiated by the stationary value of the average interface density $\langle\rho_{\rm{st}} \rangle$. These results are in agreement with the results reported for the Schelling model\cite{Gauvin_2009}, with the extra information provided by the average interface density.

\begin{figure*}
\centering
\includegraphics[width=\linewidth]{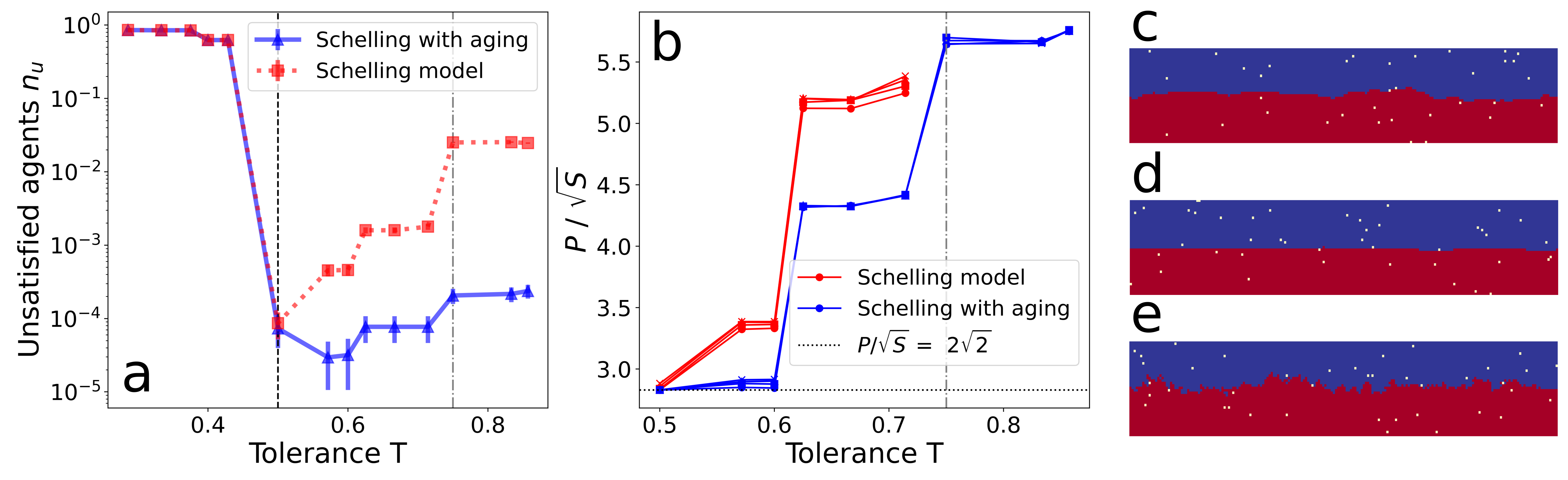}[h]
\caption{ (\textbf{a}) Fraction of unsatisfied agents $n_u$ at the stationary regime as a function of the tolerance parameter $T$. (\textbf{b}) Measure of the interface roughness between clusters of different kind of agents at the final stationary state $P/\sqrt{S}$ as a function of the tolerance parameter $T$. Different markers indicate different system sizes: $L = 40$ (circles), $60$ (squares), $80$ (triangles) and $100$ (crosses). Results are shown for both the Schelling model with and without aging. Numerical simulations are performed for $\rho_v = 0.5\%$ and averaged over $5 \cdot 10^4$ realisations. The frozen-segregated transition (dashed black line) and the segregated-mixed transition (grey dot-dashed line) are highlighted to differentiate the phases that Schelling model exhibits. (\textbf{c}) Final state interface zoom snapshot for $T = 0.57$ using the original model. (\textbf{d}) Final state interface zoom snapshot for $T = 0.57$ using the model with aging.  (\textbf{e}) Same as c for $T = 0.86$.}
\label{Fig2}
\end{figure*}

A first quite dramatic effect of including aging in the system is the disappearance of the mixed state from the phase diagram. In both metrics, the difference between the models with and without aging is clearly manifested. For low $T$ values, the frozen-segregated transition behaves similarly to the orignal model since aging has no implications as the system gets quickly frozen. Nevertheless, for high values of the tolerance $T> 0.5$, the segregated-mixed transition disappears and the segregated phase is always present. This is not an intuitive effect and one would think that aging, contributing to difficult the agents mobility, should prevent the system from forming full developed segregated clusters. However, it is just the opposite and it favors the cluster emergence. 

\subsection*{Segregated phase: final state}

To gain further insights on the differences in the system dynamics that lead to the extended segregated phase, we compute the fraction of unsatisfied agents at the stationary regime $n_u$ (see Fig. \ref{Fig2}a). This metric plays a role as a marker for the frozen-segregated transition, as shown for the 1D Schelling model \cite{Dall_Asta_2008}. The frozen phase presents a big majority of unsatisfied agents for both models. After the transition, this parameter decays to very low values in the segregated phase where a majority of agents are satisfied. In this phase, we observe a step-like increasing behaviour of the unsatisfied agents with $T$. As the tolerance grows, the number of satisfying vacancies increases and the noisy movement of satisfied agents drives the system evolution, creating eventual unsatisfied agents in the sites that they abandon or target. However, in the Schelling model, the transition to a mixed state at $T = 0.75$ inhibits the creation of clear fronts between agents of different kinds and it is also associated to a sharp increase of $n_u \simeq 0.05 $ (red squares in Fig. \ref{Fig2}a). The Schelling model with aging, on the other hand, shows a lower fraction of unsatisfied agents during all values of the tolerance above the frozen-segregated transition (blue triangles in Fig. \ref{Fig2}a). So much so, that many realizations reach $n_u = 0$ and this causes the large error bars in Fig. \ref{Fig2}a after the transition. In counter intuitive way, the introduction of aging causes a higher global satisfaction when compared with the original model in both the segregated and the mixed phases.

The creation of new unsatisfied agents at the final stationary state occurs at the interface between the segregated agent kinds. This is why we study the interface roughness as a function of the tolerance parameter. The roughness is characterized as a deviation from a flat configuration. In our system with periodic boundary conditions and a size of $L \times L$, the minimum perimeter between clusters of agent kinds is $P = 2 \, L$. To avoid the $L$ dependency, we calculate an adimensional magnitude $P/\sqrt{S}$, where $S$ is the number of agents of each kind $S = N_{\rm{agents}}/2 = L^2 \, (1 - \rho_v )/2 $. In addition, to calculate the perimeter we smooth the interface out by considering vacancies surrounded by a majority of agents of a certain kind as members of that kind. This metric $P/\sqrt{S}$ is computed starting from a flat interface as an initial condition and evolving it for $t_{\rm{max}} = 10^4$ MC steps to reach well within the stationary state. With the metric $P/\sqrt{S}$, we are able to estimate how close is the final state interface of our system to the flat interface ($P/\sqrt{S} = 2 \, \sqrt{2}$). The results show a increasing dependence of roughness with the tolerance parameter $T$ (see Fig. \ref{Fig2}b). This growth can be explained as an increase of the tolerance means that agents are satisfied with less "same-kind" neighbors. Therefore, the interface is able to be rougher keeping the agents in a satisfied state. In addition, notice that all values with different $L$ collapse so the dependence on the system size has been eliminated. 

\begin{figure*}
\centering
\includegraphics[width=0.8\linewidth]{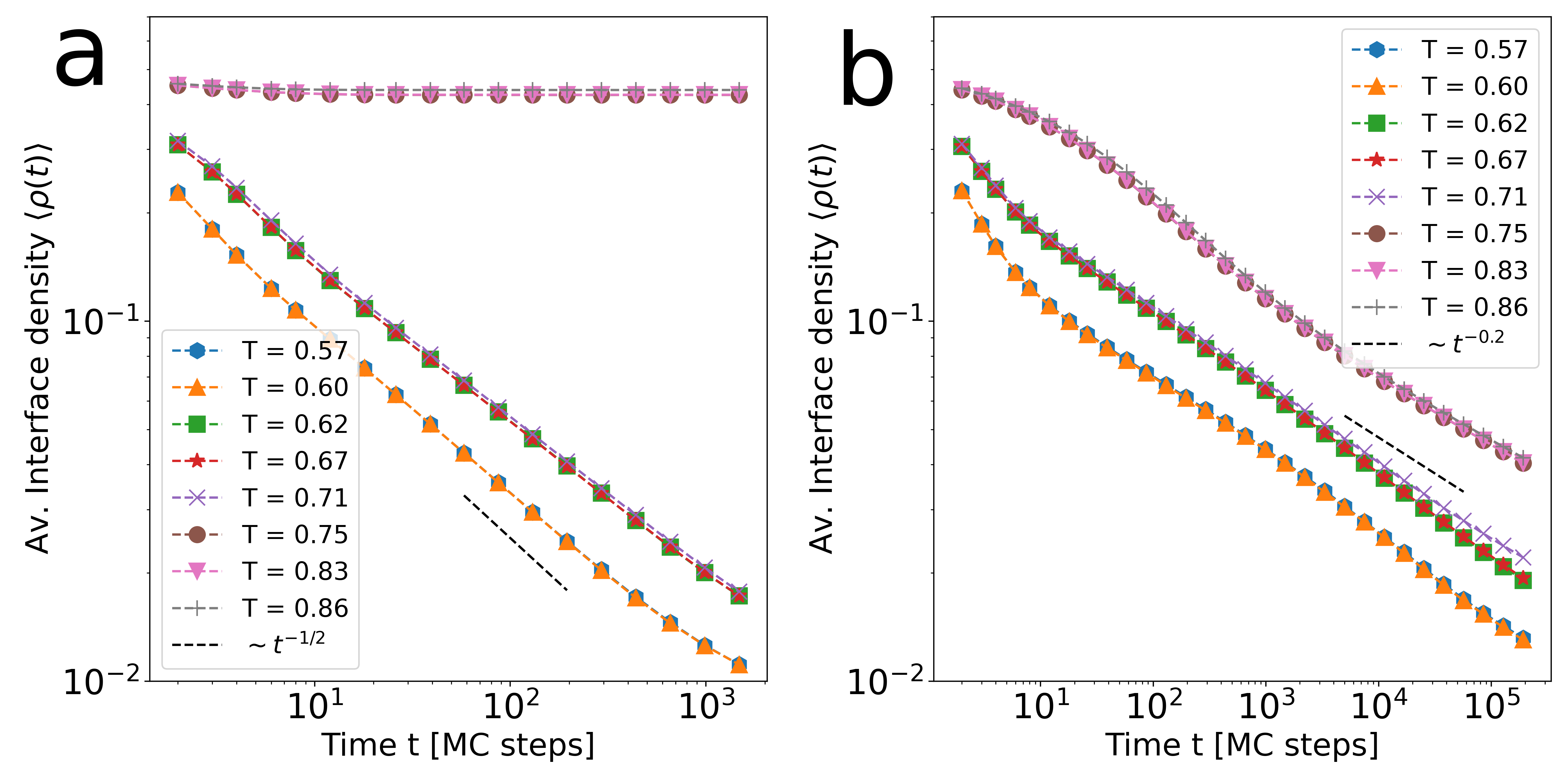} 
\caption{Average interface density $\langle \rho (t) \rangle$ as a function of time steps for different values of the tolerance parameter $T$ using the Schelling model (\textbf{a}) and the version with aging (\textbf{b}). Average performed over $5 \cdot 10^3$ realisations. Fitted power-law in a black dashed line highlighting the estimated exponent value. We set system size $L = 200$ and $\rho_v = 0.005$.}
\label{Fig3}
\end{figure*}

Comparing both models, one observes a lower interface roughness for the Schelling model with aging regardless of the value of $T$. The closest value to the flat interface occurs for the first values of $T$ after the frozen-segregated phase transition (shown in Fig. \ref{Fig2}d). In the original model, we observe higher values of $P/\sqrt{S}$ due to the noise produced by the satisfied agents' behaviour (see Fig. \ref{Fig2}c). Moreover, aging allows us to obtain a segregated phase with even larger interface roughness than the maximum observed in the original model for large values of $T$ (see Fig. \ref{Fig2}e). We remark that, when aging is introduced, agents try to join those of their own kind but are less and less prone to change location as time passes. Thus, in the Schelling model with aging, agents in the bulk of the clusters mainly do not move and those moving more often are located at the interface between agent kinds. At medium and large scales, this phenomenon leads to a ergodicity breaking in the final state dynamics.

\subsection*{Segregated phase: coarsening dynamics}

Diverse versions of the original Schelling Model exhibit different behaviors in terms of coarsening dynamics. Recent publications report a power-law like domain growth \cite{Dall_Asta_2008,Interfacial_roughening}. We monitor here the evolution of the interface density $\langle \rho (t) \rangle$, which decreases as $ \langle \rho (t) \rangle \sim t^{-\alpha}$ so the domains should grow in our model following a power-law with time. 

\begin{figure*}
\centering
\includegraphics[width=\linewidth]{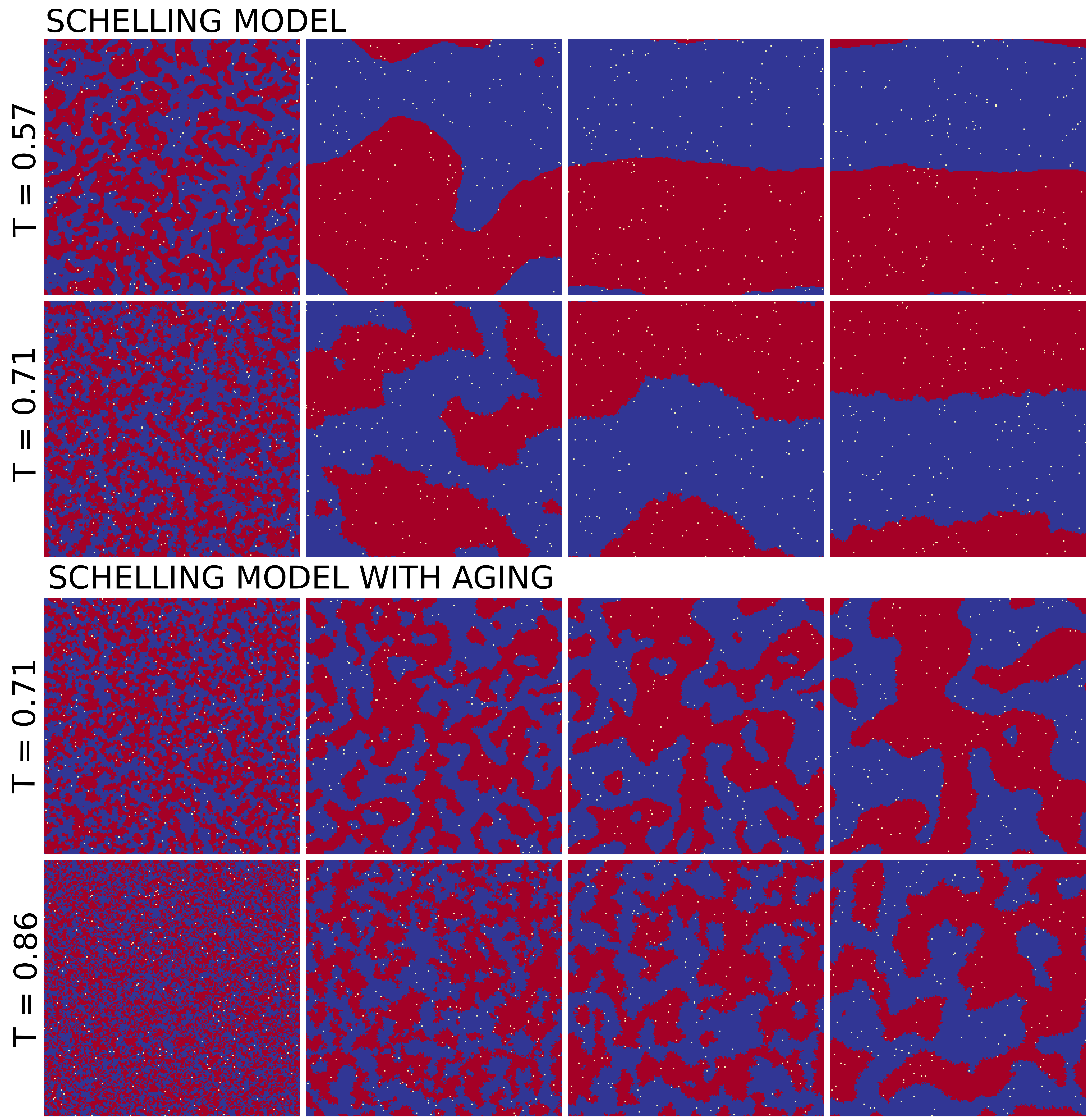} 
\caption{Coarsening towards the segregated state at two different values of $T$ for both models. Snapshots are taken for $5$, $500$, $5000$ and $50000$ time steps ordered from left to right. We set system size $L = 200$ and $\rho_v = 0.005$.}
\label{Fig4}
\end{figure*}

The coarsening process of the Schelling model at the segregated phase ($0.5 \le T < 0.75$) is displayed in Fig. \ref{Fig3}a and Fig. \ref{Fig4}. We find that the average interface density follows a power-law decay with an exponent $\alpha \simeq 0.5$ for the limit of small vacancy density $\rho_v \to 0$, in agreement with the value reported for close variants of the Schelling model \cite{Dall_Asta_2008}. This exponent value is curious, since the coarsening in the presence of a conserved quantity (but with local interactions) exhibits an exponent $\alpha = 1/3$ \cite{Maxi}. Nevertheless, the interactions in this model are not local and the coarsening exponent is more similar to the one in systems with non conserved order-parameter ($\alpha = 1/2$). Fig. \ref{Fig3}a shows as well how coarsening changes with the tolerance parameter. Even though the exponent $\alpha$ does not depend on $T$, we observe a certain delay when increasing $T$ from $0.6$ to $0.62$. In the system evolution of Fig. \ref{Fig4}, one can see how the behaviour of the satisfied agents for higher tolerance values is translated into rougher interfaces, causing such delay. For $T > 0.75$, the system exhibits a transition towards a mixed state where the interface density fluctuates around $\rho = 0.5$ indicating that the state is constantly disordered.

The Schelling model with aging shows very different behaviour (Fig. \ref{Fig3}b). As predicted by the phase diagram, the average interface density exhibits a power-law decay with time for all values of the tolerance $T$ after the frozen-segregated transition. Still, the decay is slower than for Schelling model, with $\langle \rho (t)\rangle \; \sim \; t^{-0.2}$. A mechanism that could be behind this behavior is that the model with aging counts with more satisfied agents than the original model and their probability to move becomes lower as time goes by. Moreover, satisfied agents inside a cluster will not move and the dynamics in the model takes place at the interface. It is, therefore, more difficult for separated clusters to collide and merge, an effect that slows down the decay of the interface density. The persistence of small clusters becomes clear when the snapshots evolution is compared for both models at the same tolerance value $T = 0.71$ (see Fig. \ref{Fig4}). Moreover, while for the original model the initial clustering for $t = 500$ steps does not determine the final state, in the case with aging the bigger clusters present at the beginning of the evolution are the ones that keep growing determining the shape of the system configuration after $50000$ time steps. This is a dynamical effect, because the system in both cases tends to a final configuration with 2-clusters.

In the case of the Schelling model with aging, we observe an early cross-over in the dynamics (Fig. \ref{Fig3}b). For $T < 0.75$, the coarsening starts with an initial decay of $\langle \rho (t)\rangle$ faster than $t^{-0.2}$. This occurs because in this regime it is necessary some time for the aging effects to become relevant and before it the system behaves as in the original model. Similarly, for $T \ge 0.75$, $\langle \rho (t)\rangle$ decays slowly for a moment before reaching the power-law behavior for large $t$ values.  Confirming this scenario, Fig. \ref{Fig4} shows that for $T = 0.86$, the system starts evolving similarly to a mixed state until some clusters are created. At this moment, aging prevents the clusters desegregation leading the system very slowly to a coarsening dynamics and, eventually, to a fully segregated state. 

Regarding the relaxation time to the final state, we see in Fig. \ref{Fig4} how for $T = 0.71$, the stationary state of the Schelling model is reached after approximately $t = 5000$ time steps. In contrast, the version with aging needs much more than $50000$ steps to attain it. This highlights the important temporal  difference between both models in terms of domain growth dynamics, which strongly increases the computational cost of the study of the stationary state of the model with aging. We have been thus able to study only medium and small system sizes in this final regime (see videos included as Supplementary Information S1 and S2).

The dynamics studied thus far are preformed considering the limit $\rho_v \to 0$, but the analysis can be extended to higher vacancy densities. For the particular case of high $\rho_v$ and low $T$, aging leads to the formation of a vacancy cluster at the interface between domains (see details in Supplementary information S3).

\subsection*{Aging breaks the asymptotic time-translational invariance}

\smallskip

The very slow dynamics of the model with aging establishes a parallelism with glassy systems. Here, we explore further this similarity by considering the presence or absence of time translational invariance (TTI) in the model dynamics. For this, we start by defining the two-time autocorrelation function $C(\tau,t_{\rm{w}})$ as

\begin{equation}
    C(\tau,t_{\rm{w}}) = \left\langle \frac{1}{M} \, \sum_{i = 1}^{N}  \sigma_i (t_{\rm{w}} + \tau) \,  \sigma_i(t_{\rm{w}}) \right\rangle ,
\end{equation}
where $N$ is the system size,  $\langle . \rangle$ refers to averages over realizations, $t_{\rm{w}}$ is the waiting time to start the autocorrelation measurements, $\tau$ a time interval after $t_{\rm{w}}$ and $M$ is a normalization factor defined as
\begin{equation}
M =  \sum_{i = 1}^{N}  (\sigma_i (t_{\rm{w}} + \tau) \, \sigma_i(t_{\rm{w}}))^2 . 
\end{equation}
Note that $M$ is calculated in each realization and the average is only taken over the final correlation. 

\begin{figure*}
\centering
\includegraphics[width=\linewidth]{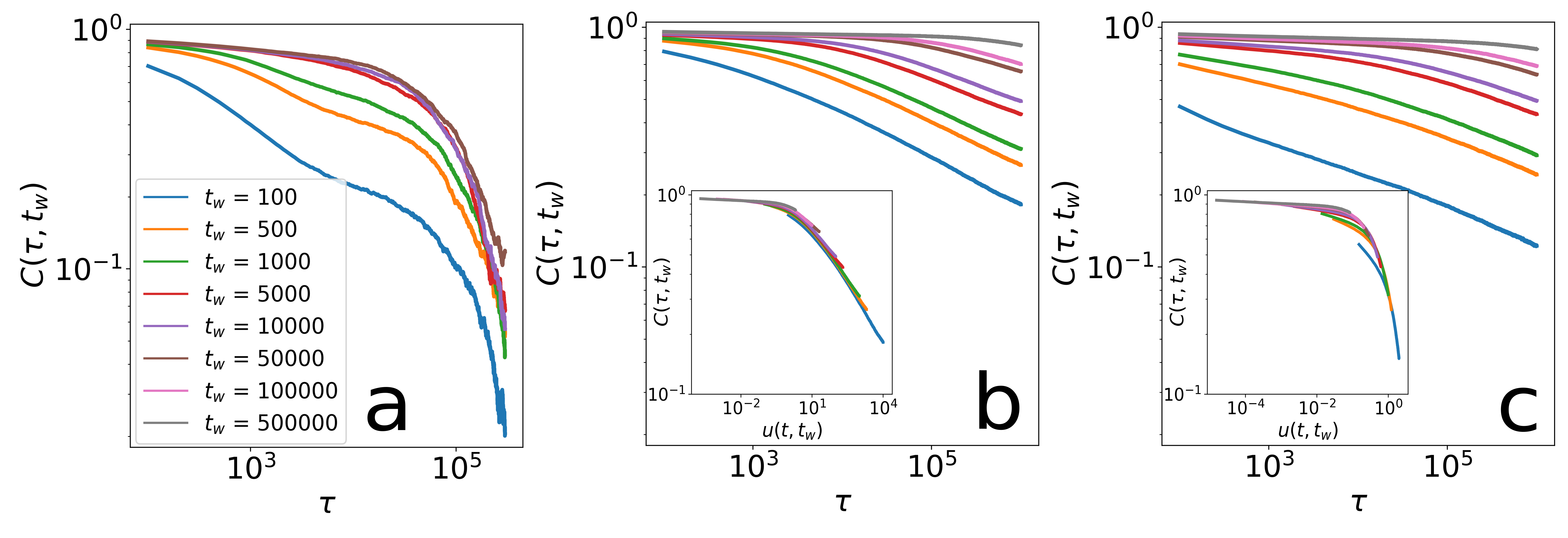} 
\caption{Two-times autocorrelation $C (\tau,t_{\rm{w}})$ as a function of the time period passed since the waiting time $t_{\rm{w}}$. First, the autocorrelation is shown for the Schelling model at $T = 0.71$ in \textbf{a}, and for the version with aging at $T = 0.71$ in \textbf{b} and $T = 0.86$ in \textbf{c}. The insets are result of the collapse using $u(\tau,t_w) = \tau/t_w$ (\textbf{b}) and $u(\tau,t_w) = \log(\tau+t_w)/\log(t_w) - 1$ (\textbf{c}). The curves correspond to different values of the waiting time $t_{\rm{w}}$. Calculations performed on a $80 \times 80$ lattice averaged over $5 \cdot 10^{4}$ realisations.}
\label{Fig5}
\end{figure*}

The autocorrelation function is displayed for the Schelling model with $T = 0.75$ in Fig. \ref{Fig5}a. We observe the curves decreasing with $\tau$ as expected, and that after a characteristic time period ($t_{\rm{w}}^* \approx 5000$ for a system size of $80\times 80$) they collapse into a single curve. This is the regime in which the dynamics becomes TTI, implying that the autoccorrelation function does not depend any more on the waiting time, $C(\tau,t_{\rm{w}}) = C(\tau)$ for $ t_{\rm{w}} > t_{\rm{w}}^{*}$. 

In the case of the Schelling model with aging, the dynamics show some different features (Figs. \ref{Fig5}b and \ref{Fig5}c). First, the autocorrelation functions decay slower with $\tau$ in all the cases, which is connected to the long-lived small clusters mentioned previously. We do not find in the simulations any value of $t_{\rm{w}}^*$ for the systems to fall into a TTI regime. Not only that, but a scaling relation including both $\tau$ and $t_{\rm{w}}$ can be applied to collapse the autocorrelation curves (see insets Figs. \ref{Fig5}b and \ref{Fig5}c). This behavior is similar to glassy systems \cite{spinglassbook}. In this type of dynamics, as for spin glasses, a final stationary state is not attainable in the thermodynamic limit and it is possible to decompose the autocorrelation function into an equilibrium part and an "aging" part (aging in the sense of non equilibrium dynamics in spin glasses)  \cite{spinglassbook,Heisemberg}:

\begin{equation}
    C (\tau, t_{\rm{w}}) \simeq C_{\rm{eq}}(\tau) \; C_{\rm{aging}} u(\tau,t_w) = C_{\rm{eq}}(\tau) \; C_{\rm{aging}} \left( \frac{h (\tau) }{h(t_{\rm{w}})} \right),
\end{equation}

where $C_{\rm{eq}}$ describes the fast relaxation of the system components within each domain (TTI term), $C_{\rm{aging}}$ is an scaling function and $u(\tau,t_w)$ is a normalization factor which, in some cases, can be written as the quotient of an unknown function $h(t)$ at the two times $\tau$ and $t_{\rm{w}}$. This function $h(t)$ is known to be related with the dynamical correlation length  \cite{Heisemberg,8Heisemberg}. In our case, we use $h(t) = t$ to scale the results in Fig. \ref{Fig5}b (see inset). This scaling is valid for values of $T \in [0.5,0.75)$. Nevertheless, higher values of $T$ do not hold a linear scaling and we need to turn to other functional forms as the normalization factor $u(\tau,t_w) =  \log(\tau+t_w)/\log(t_w) - 1$ used in  Fig. \ref{Fig5}c. This indicates that for $T > 0.75$, the dynamical correlation length evolves in a different and slower way.

\smallskip

\section*{Summary and discussion}

\smallskip

We have studied the effect of aging on a stochastic threshold model which combines long-range mobility with local short-range interactions. Specifically, taking as basis the noisy constrained Schelling model, we assign to the agents an internal clock counting the time spent in the same satisfying location. The probability of changing state decreases then inversely proportional to this time. Therefore,  older satisfied agents are less prone to update resident locations. The original model displays a transition between a segregated phase and a mixed one as the tolerance control parameter $T$ increases. This transition disappears when aging is introduced into the system, the mixed phase is replaced by a segregated phase even for high values of the tolerance parameter $T$. As a result, the model with aging presents a higher global satisfaction than without this effect for all values of the tolerance. 

On the dynamical perspective, the relaxation towards the segregated phase features a coarsening phenomena characterized by a power-law decay of the average interface density with time $\langle \rho \rangle \sim t^{-\alpha}$. For the original model in the limit of low vacancy density, the exponent is around $\alpha = 1/2$. This exponent is also reported in other variants of the Schelling model \cite{Dall_Asta_2008,Interfacial_roughening}. Aging gives rise to long-lived small clusters and a slower coarsening, reducing the exponent to $\alpha \simeq 0.2$. We investigated the autocorrelation functions in the segregated phase and found that aging breaks the asymptotic time-translational invariance of the dynamics. This result, along with a nontrivial scaling of the autocorrelation functions, establish close similarities between glassy dynamics and our Schelling model with aging for high values of the tolerance parameter. 

As for the implications of our results from a social perspective, we must note that the fact that aging favors segregation, inhibiting the segregation-mixed phase transition, is rather counter-intuitive, but gives support to the argument that segregation is a stochastically stable state and may prevail in an all-integrationist world \cite{Zhang}. Our model predicts that the appearance of segregation even for tolerance  values close to one. Additionally, the model relaxation time multiplies manifold, which implies that if aging is present the natural state of this system seems to be generically out of equilibrium.   


%

\section*{Acknowledgements}

Partial funding is acknowledged from the project PACSS (RTI2018-093732-B-C21, RTI2018-093732-B-C22) of the MCIN/AEI/10.13039/501100011033/ and by EU through FEDER funds (A way to make Europe), and also from the Maria de Maeztu program MDM-2017-0711 of the MCIN/AEI/10.13039/501100011033/.

\section*{Author contributions statement}

D.A., M.S.M. and J.J.R. conceived and designed the research. All the authors analyzed the results.  D.A. performed the simulations. All authors contributed to scientific discussions and to writing the paper.

\section*{Additional information}

The authors declare no competing interests.

\end{document}